# Why Ammoniated Lithium Borohydrides Liquefy and Resolidify?

Qian Wang[1 †], Zixin Xu[2 †], Ryuhei Sato[3], Yiming Lu[1], Hiroki Miyaoka[4], Takayuki Ichikawa[2], Eric Jianfeng Cheng[1], Shin-ichi Orimo[1,5], Fangqin Guo[2, *], Hao Li[1, *]

[1] Advanced Institute for Materials Research (WPI-AIMR), Tohoku University, Sendai 980–8577, Japan

[2] Graduate School of Advanced Science and Engineering, Hiroshima University, 1-4-1 Kagamiyama, Higashi-Hiroshima 739-8527, Japan

[3] Department of Materials Engineering, The University of Tokyo, 7-3-1 Hongo, Bunkyo-ku, Tokyo 113-8656, Japan

[4] Core Facility Management Center, Hiroshima University, 1-3-1 Kagamiyama, Higashi-Hiroshima, 739-8530, Japan

[5] Institute for Materials Research (IMR), Tohoku University, Sendai 980–8577, Japan

[†] Equal contributions

* Corresponding authors, E-mails: fang-qin-guo@hiroshima-u.ac.jp (F.G.), li.hao.b8@tohoku.ac.jp (H.L.)

## Abstract

Ammonia ($NH_3$) absorption drives $LiBH_4 \cdot xNH_3$ through a re-entrant "solid–liquid–solid" transition: **$LiBH_4 \cdot NH_3$ is a well-defined solid ammoniate, compositions near $LiBH_4 \cdot 2NH_3$ are liquid-like or partially liquefied, whereas $LiBH_4 \cdot 3NH_3$ returns to a more rigid non-liquid ammoniate state.** However, the microscopic origin of this unintuitive response remains a long-lasting mystery. Here, we uncover its mechanism. Cross-database analysis identifies borohydrides as a particularly state-diverse and composition-responsive material family. Structure prediction and *ab initio* molecular simulations reveal that increasing $NH_3$ loading increases the direct Li–N coordination number while progressively decreasing $BH_4^-$-associated contacts in the local Li environment. Near $x \approx 2$, these contributions are most balanced among the simulated compositions, and the sampled Li–N/N···B coordination landscape is broadest. Further ammoniation produces Li–N-dominant coordination and slower $BH_4^-/NH_3$ contact renewal, accompanying recovery of a more rigid ammoniate state. Pressure–composition isotherm, $^1H$ and $^{11}B$ nuclear magnetic resonance, and Raman measurements support this non-monotonic state evolution and associated $BH_4^-/NH_3$ reorganization. These findings transform ammonia-induced liquefaction from an empirical phase anomaly into a competition between native-network disruption, mixed-coordination frustration, and ligand-built network reconstruction, providing a framework for chemically switching between transport-favouring fluidity and stability-favouring rigidity in hydrogen-rich materials.

## INTRODUCTION

Complex hydrides constitute a versatile family of hydrogen-rich materials relevant to hydrogen and ammonia storage, electrochemical energy storage, and related energy-conversion processes.[1, 2, 3] Their appeal arises from their high gravimetric and volumetric hydrogen densities and from the structural and dynamical flexibility of complex anions.[4] However, their performance is often constrained by slow transport through rigid condensed phases, insufficient reactant contact, and interfacial resistance.[5, 6] Physical state is therefore an important yet underdeveloped design variable: liquid-like phases can enhance mass transport and wetting, whereas solids provide dimensional and mechanical stability.[3] Hydrogen-rich liquids formed from complex hydrides further demonstrate the potential of deliberately controlling hydride phase state.[7] Controlled switching between fluidity and rigidity could thus reconcile fast transport with structural robustness, but the microscopic principles governing such switching remain poorly understood.

Borohydrides, containing the tetrahydridoborate anion ($BH_4^-$), are particularly suitable for examining this problem because the $BH_4^-$ anion combines high hydrogen content with rotational dynamics and adaptable coordination. $LiBH_4$ has a high theoretical hydrogen capacity of 18.5 wt%,[8, 9] while borohydride- and *closo*-borate-based electrolytes can reach $Li^+$ or $Na^+$ conductivities of $10^{-3}$–$10^{-2}$ S $cm^{-1}$ when anion rotation, substitution, disordering, or molecular-complex formation is activated.[10, 11, 12] Borohydrides can also reversibly absorb $NH_3$ at high density,[13, 14] while $LiBH_4$ and $NaBH_4$ can form low-viscosity liquid ammoniate phases under certain conditions.[15] In ammine metal borohydrides, $NH_3$ coordinates directly to the metal through the N lone pair, whereas $BH_4^-$ can act as a terminal or bridging ligand or, at high ammine content, as a counterion.[5, 16, 17, 18, 19] Increasing $NH_3$ loading therefore changes direct metal–N coordination and $BH_4^-$-associated contacts, with consequences for molecular mobility and material state.

$LiBH_4 \cdot xNH_3$ provides a striking example of composition-driven re-entrant behavior.[13, 20, 21, 22] Notably, their physical states change **non-monotonically** with $NH_3$ content: **$LiBH_4 \cdot NH_3$ is a** **well-defined solid ammoniate, the compositions near $LiBH_4{\cdot}2NH_3$ are liquid-like or partially liquefied, whereas $LiBH_4{\cdot}3NH_3$ returns to a more rigid ammoniate state.**[13, 15, 23] Early pressure-composition-temperature (PCT) measurements revealed a strongly composition-dependent phase diagram, with distinct phase fields around $x \approx 0.5$, 1, 2, 3, and 4 and lower liquidus temperatures across several intervening composition ranges.[23] The structure and melting behavior of $LiBH_4{\cdot}0.5NH_3$ provide a well-defined low-$NH_3$ reference state,[12] while compositions within reported two-phase regions represent phase coexistence rather than unique single phases. Nevertheless, these macroscopic phase relationships do not explain why initial $NH_3$ uptake disrupts the solid structure whereas further ammoniation restores rigidity. The intermediate region around overall $x \approx 2$ may involve eutectic melting or solid–liquid coexistence rather than a unique homogeneous $LiBH_4{\cdot}2NH_3$ phase, complicating direct crystallographic determination of its local structure.[23]

Here, we investigate the structural and dynamical origin of $NH_3$-dependent "solid–liquid–solid" evolution in $LiBH_4{\cdot}xNH_3$ ($x$ = 0.5, 1, 2, 3, and 4). Artificial intelligence (AI) agent-assisted cross-database analysis is combined with structure prediction, finite-temperature molecular dynamics, enhanced sampling, and experimental measurements. To the best of our knowledge, this work provides the first mechanistic explanation of this long-standing phenomenon. Increasing $NH_3$ loading increases direct Li–N coordination and decreases neighboring $BH_4^-$ contributions; the fixed $x = 2$ reference model exhibits the most balanced contributions. Further ammoniation produces Li–N-dominant coordination and a more persistent $BH_4^-/NH_3$ network, accompanying the return to a rigid state. These results identify the interplay between local coordination and contact persistence as the microscopic origin of the re-entrant transition and provide a potentially transferable framework for understanding transient fluidity in other complex hydrides.

## Results and Discussion

### Cross-database analysis of ammonia-induced state diversity

Due to the limited available $NH_3$-storage dataset for hydrides, comprising approximately 12

studies[15, 23, 24, 25, 26, 27, 28, 29, 30, 31, 32, 33] and 28 materials in the present collection, related hydride records from our recently developed Digital Battery platform (*DigBat*)[34] and Digital Hydrogen platform (*DigHyd*)[35] were also included as complementary references rather than as direct measures of material state (**Fig. 1a**). *DigBat* contains ~2,000 experimental data entries of >200 hydride electrolytes, while *DigHyd* comprises >30,000 hydride-related data entries. Within the ammonia storage dataset, liquid and solid–liquid coexistence states are concentrated primarily in ammonia borane and borohydrides, whereas $NH_3$-loaded metal halides are reported predominantly as solids (**Fig. 1b**). Borohydrides exhibit the broadest state diversity, spanning solid, liquid, and composition-dependent or transitional states. This contrast demonstrates that $NH_3$ incorporation alone is insufficient to induce liquefaction. **Instead, the resulting physical state likely depends on how $NH_3$ disrupts the original intermolecular or anion-supported network and whether a new persistent coordination network is subsequently established.**

*DigBat* and *DigHyd* were then used to examine whether related $BH_4^-/NH_3$ materials show composition- and cation-dependent responses (**Fig. 1c, the references are listed in Supplementary Table 1**). In *DigBat*, the ionic conductivities of solid electrolytes were compared at 303 K after Arrhenius normalization using the reported activation energies. The $LiBH_4$-based series is limited to $NH_3/BH_4$ = 0, 0.5, and 1, corresponding to $LiBH_4$, $LiBH_4{\cdot}0.5NH_3$, and $LiBH_4{\cdot}NH_3$, and its conductivity increases with increasing $NH_3/BH_4$ ratio. Higher-loading $LiBH_4{\cdot}2NH_3$, $LiBH_4{\cdot}3NH_3$, and $LiBH_4{\cdot}4NH_3$ are not included in these solid electrolyte structural/property records because they correspond to liquid-like, partially liquefied, or structurally unresolved regimes. In contrast to the monotonic Li-based trend, the $Mg(BH_4)_2$-based series shows a non-monotonic response, reaching a maximum conductivity of approximately $5\times10^{-7}$ S $cm^{-1}$ near $NH_3/BH_4$ = 0.9 and then decreasing at higher $NH_3$ content.[36]

In *DigHyd*, the hydrogen desorption capacities for Mg- and V-containing borohydrides and theoretical total hydrogen contents for Mn-containing borohydrides tend to increase with $NH_3/BH_4$ ratio under different conditions. Together with the ammonia storage records, **these trends identify the $NH_3/BH_4$ ratio as a key compositional variable governing material-state and property responses, while showing that the direction and magnitude of these responses depend on cation identity and the measured property.** AI agent-assisted workflow organized composition- and cation-dependent evidence across the three datasets for comparison (**Supplementary Figs. 1–4** and **Supplementary Videos 1-2**). These comparisons motivated a focused analysis of the $Li^+$ coordination environment in $LiBH_4{\cdot}xNH_3$, where the cation identity remains fixed while the $NH_3/BH_4$ ratio varies.

Experimentally reported structures were used for $LiBH_4{\cdot}0.5NH_3$ and $LiBH_4{\cdot}NH_3$, whereas candidate structures for $LiBH_4{\cdot}2NH_3$, $LiBH_4{\cdot}3NH_3$, and $LiBH_4{\cdot}4NH_3$ were obtained from USPEX[37] structure searches based on genetic algorithm (**Supplementary Figs. 5-9 and Supplementary Table 2**). The predicted structures are computational reference models, not experimentally established ground-state structures. The V/Z trend provides a structural-model consistency check rather than same-composition experimental validation (**Supplementary Fig. 5e**). Representative structures show a progressive change in the Li coordination environment (**Fig. 1d**): direct Li–N coordination is already present at $x = 0.5$; with increasing $x$, Li–N contributions increase and $BH_4^-$-associated contributions decrease, becoming most comparable near $x = 2$. This qualitative progression was subsequently quantified through Li-shell composition, coordination exchange, and $BH_4^-/NH_3$ network connectivity.

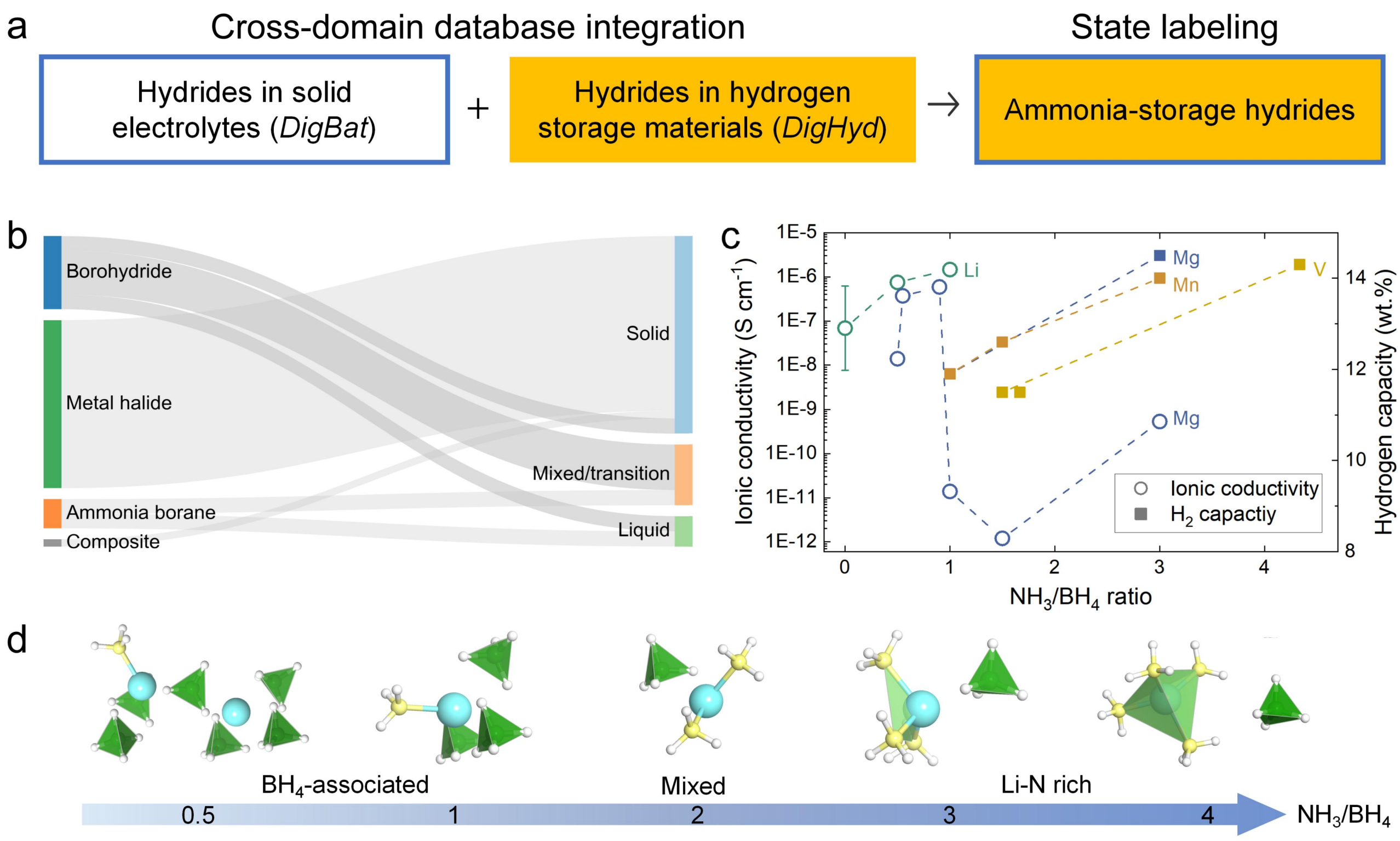

**Fig. 1. Cross-database patterns in ammonia-induced state and property evolution. a**, Complementary data domains for ammonia-induced state anomalies. **b**, Reported physical-state distribution of ammonia-storage borohydrides, metal halides, ammonia boranes, and composites. **c**, Ionic conductivity and hydrogen capacity as functions of $NH_3/BH_4$ ratio for related borohydride materials from *DigBat* and *DigHyd* databases, respectively; original literature sources and values are given in **Supplementary Table 1**. Conductivities are normalized to 303 K; hydrogen-storage data include experimental $H_2$-release capacities and literature-reported theoretical total hydrogen contents for the Mn compounds. **d**, Representative Li coordination environments in $LiBH_4{\cdot}xNH_3$ with increasing $x$ loading; cyan, green, yellow, and white spheres represent Li, B, N, and H atoms, respectively.

## Li coordination-shell switching defines a mixed-shell regime

Before examining the finite-temperature coordination dynamics, the relative 0 K stability of the candidate structures was evaluated. The corresponding 0 K electronic-energy convex hull places the $x$ = 1, 3, and 4 reference compositions on the hull, whereas $x$ = 0.5 and $x$ = 2 lie 1.34 and 11.49 meV atom$^{-1}$ above it, respectively (**Supplementary Fig. 10**).

Unbiased *ab initio* molecular dynamics simulations were used to quantify how $NH_3$ loading changes the local Li environment in $LiBH_4{\cdot}xNH_3$ after structural equilibration. Li–N coordination was calculated from N atoms within a common first-shell cutoff, whereas Li–B distances were used as a one-center-per-$BH_4^-$ molecular-neighbor descriptor (**Supplementary Fig. 11a, b**). All values are ensemble averages over Li sites and production frames and are therefore not restricted to integers. Conventional Li – H($BH_4$) RDFs and coordination numbers were additionally evaluated as an independent measure of Li–$BH_4$ association (**Supplementary Fig. 11c–d**).

RDF analysis shows a clear substitution of the Li coordination shell with increasing $NH_3$ loading. Li–N correlations exhibit a pronounced first-shell peak near 2 Å for all $LiBH_4 \cdot xNH_3$ compositions, and the first-shell contribution increases with $x$ (**Fig. 2a**). In contrast, the Li–B first-shell correlations decrease and broaden with increasing $NH_3$ loading (**Fig. 2b**). The corresponding neighbor statistics confirm this opposite trend: CN(Li–N) increases, whereas CN(Li–B) decreases (**Fig. 2c**). The conventional Li–H($BH_4$) analysis independently confirms this trend: CN[Li–H($BH_4$)] decreases from 7.52 at $x$ = 0.5 to 2.00 at $x$ = 4 using a 2.87 Å cutoff (**Supplementary Fig. 11c–d**). The two contributions are most comparable at $x$ = 2, whereas Li–N coordination becomes dominant by $x$ = 3. $LiBH_4 \cdot 2NH_3$ therefore represents a mixed-coordination regime near the crossover from $BH_4^-$-rich to $NH_3$-rich Li environments.

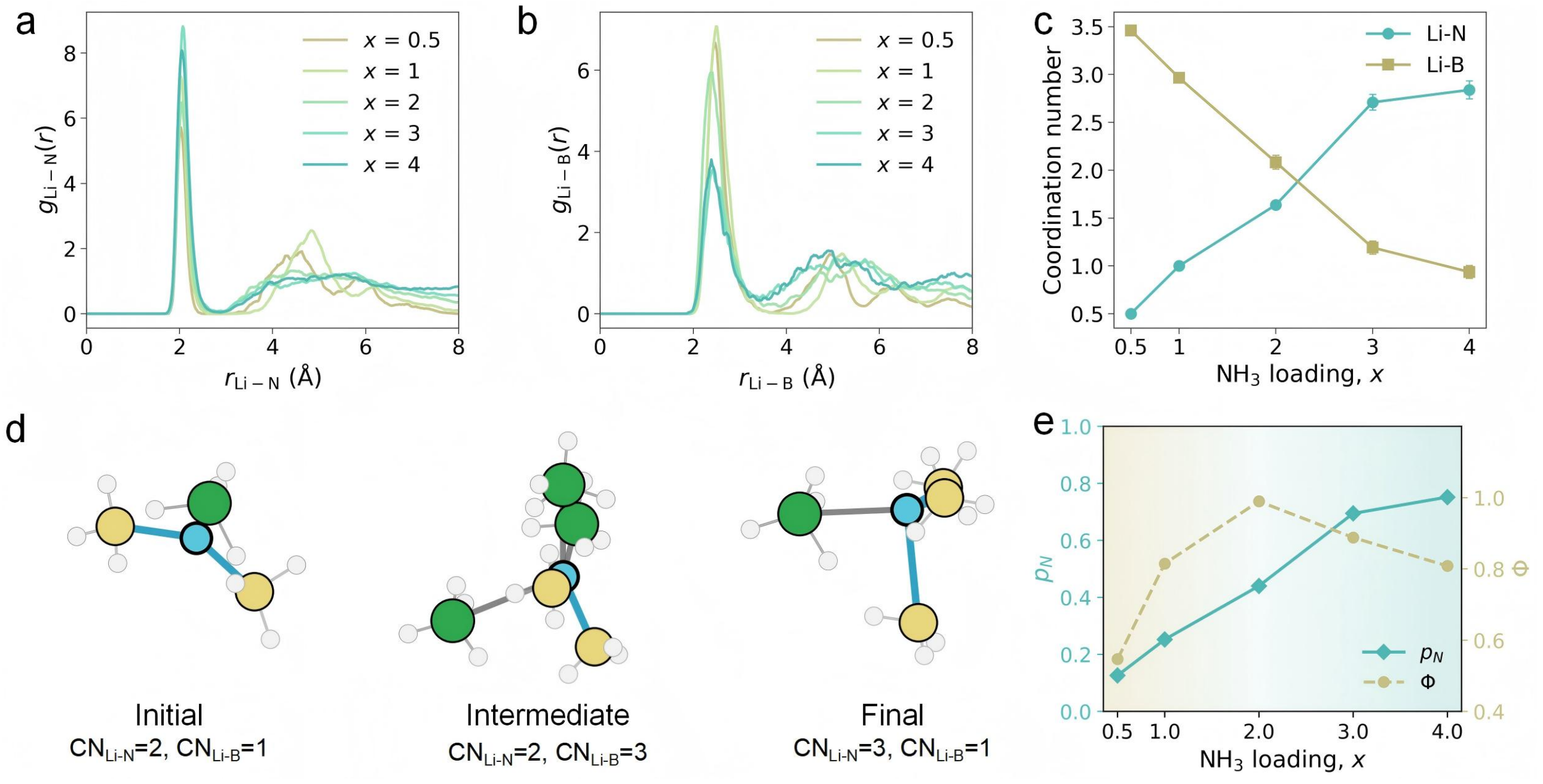

**Fig. 2. Li coordination-shell switching in $LiBH_4 \cdot xNH_3$. a**, **b**, Li–N (**a**) and Li–B (**b**) radial distribution functions (RDFs) for different $x$. **c**, First-shell Li–N coordination number and Li–B molecular-neighbor count as a function of $NH_3$ loading. **d**, Representative local configurations from an unbiased *ab initio* molecular dynamics trajectory of $LiBH_4 \cdot 2NH_3$; cyan, green, yellow, and white spheres denote Li, B, N, and H atoms, respectively. **e**, $NH_3$ fraction in the Li coordination shell, $p_N$, and normalized shell-mixing descriptor, Φ, as a function of $NH_3$ loading.

Representative configurations from the $LiBH_4 \cdot 2NH_3$ trajectory illustrate this mixed-shell regime (**Fig. 2d**). The joint Li–N/Li–B coordination distributions show the corresponding composition-dependent shift from $BH_4^-$-rich to $NH_3$-rich environments, with $LiBH_4 \cdot 2NH_3$ occupying the most pronounced mixed-shell region (**Supplementary Fig. 12**).

To quantify Li-shell composition and mixing, we defined the $NH_3$ fraction ($p_N$), and the normalized binary Shannon function of the Li-shell composition (Φ), hereafter referred to as the shell-mixing descriptor:[38]

$$p_N = \frac{CN_{Li-N}}{CN_{Li-B}+CN_{Li-N}},\ \Phi = -\frac{p_N \ln(p_N) + (1-p_N)\ln(1-p_N)}{ln2}.$$

Here, Φ is a dimensionless information-theoretic descriptor of shell composition and has no direct thermodynamic interpretation. Φ = 0 when only one component is present and Φ = 1 for equal Li–N and Li–B contributions. As shown in **Fig. 2e**, $p_N$ increases monotonically with $x$, reflecting an increasing proportion of directly coordinated $NH_3$ relative to neighboring $BH_4^-$ units in the Li-centered local environment. By contrast, Φ varies non-monotonically and reaches its maximum at $x$ = 2. Therefore, $LiBH_4 \cdot 2NH_3$ exhibits the highest Li-shell mixing among the compositions examined. This mixed-shell regime motivates the following *ab initio* metadynamics (MetaD) analysis of the coordination landscape.

**Metadynamics (MetaD) simulations reveal non-monotonic broadening of the coordination landscape**

*Ab initio* MetaD[39] was used to compare the coupled local coordination space sampled by the composition-fixed models. The Li–N coordination variable measures direct $NH_3$ coordination to $Li^+$, whereas N···B is a smooth molecular-pair descriptor of $NH_3$–$BH_4^-$ association. The Li–B molecular-neighbor descriptor decreases as the Li–N contribution increases (**Supplementary Fig. 13**); therefore, the N···B variable was selected to provide complementary information on network reorganization. At $x = 1$, the bias-derived coordination landscape is confined to a relatively localized region at high Li–N coordination and strong N···B network contact, consistent with a comparatively well-defined coupling between $NH_3$ coordination to $Li^+$ and the surrounding $BH_4^-$ framework (**Fig. 3a**). At $x = 2$, the sampled region expands markedly, particularly along the Li–N coordination dimension and covers the broadest coupled Li–N/N···B coordination region (**Fig. 3b**). Although $x = 4$ samples a comparable N···B range, it remains confined to a narrower Li–N-dominant region (**Supplementary Fig. 14**). Internal pressures were comparable across compositions in both the unbiased and MetaD simulations, with no anomaly at $x = 2$, supporting their use for comparative analysis of local coordination dynamics (**Supplementary Table 3**).

The time-dependent position of the dominant basin provides a complementary view of coordination-space migration during MetaD sampling (**Fig. 3c**). Low- and high-$NH_3$ compositions remain confined to narrower regions of the Li–N/N···B plane, whereas $LiBH_4{\cdot}2NH_3$ shows a larger displacement, indicating easier access to adjacent cation-shell and anion-framework configurations. The corresponding time-resolved collective-variable traces are shown in **Supplementary Fig. 15**.

The expanded landscape was quantified by the 95% sampled region ($A_{95}$), and by the fluctuation of the N···B network-contact coordinate (**Fig. 3d**). $A_{95}$ reaches a maximum at $x = 2$, showing that the intermediate composition explores the broadest coordination region within the present sampling window. N···B fluctuations increase with $NH_3$ loading and remain substantial at higher $x$, indicating continued reorganization of the $BH_4^-/NH_3$ network. $LiBH_4{\cdot}2NH_3$ is therefore distinguished not by network fluctuations alone, but by their coexistence with the broadest coupled Li-shell/network landscape. Both metrics approach stable values during the late stage of the simulations (**Supplementary Fig. 16**).

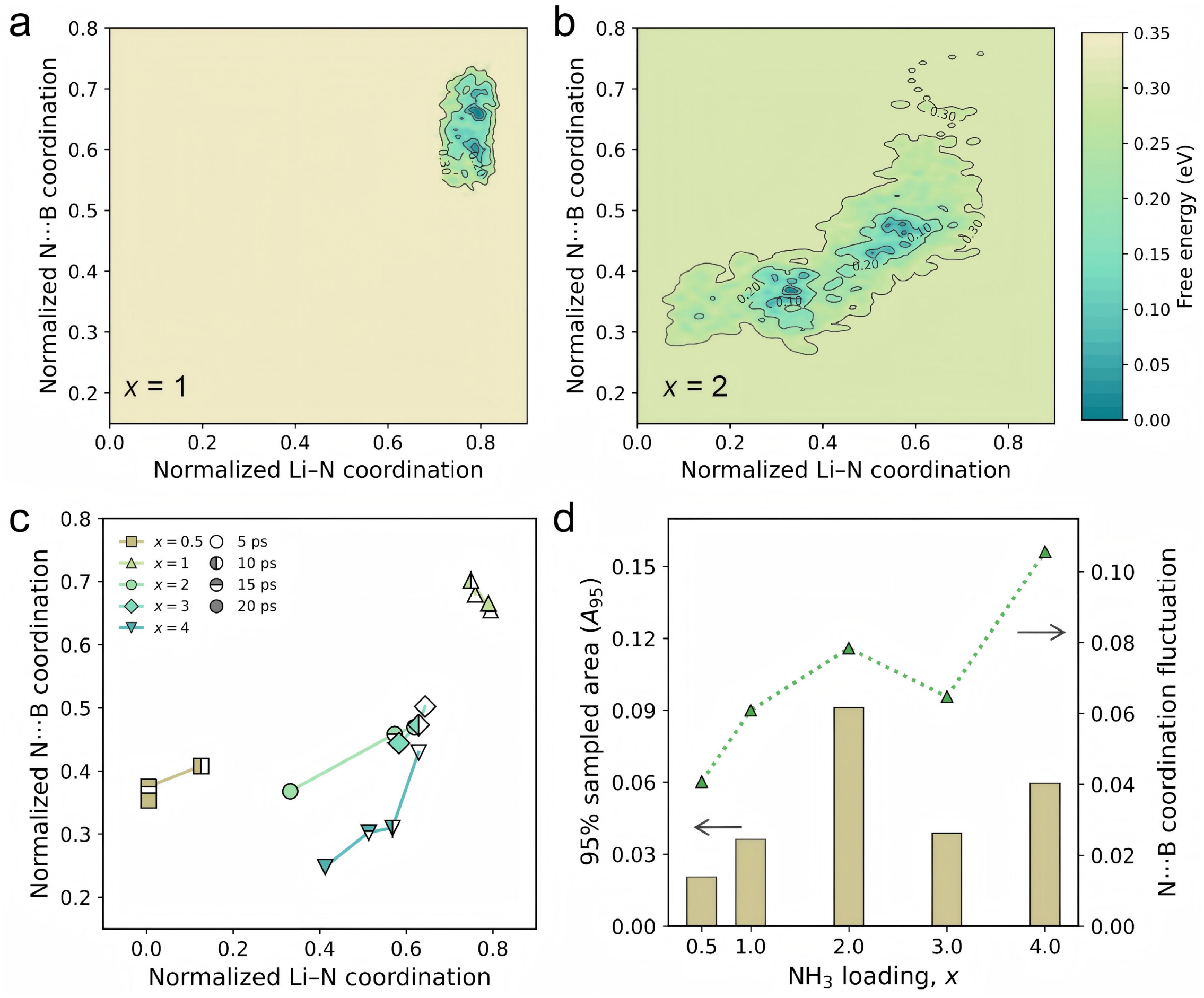

**Fig. 3. MetaD exploration of coordination space in $LiBH_4 \cdot xNH_3$. a**, **b**, Bias-derived coordination landscapes at $x$ = 1 (**a**) and $x$ = 2 (**b**), projected onto the normalized Li–N coordination and N···B network-contact variables. **c**, Time evolution of the dominant-basin positions during MetaD sampling. **d**, Composition dependence of the area containing 95% of the sampled collective-variable points ($A_{95}$) and N···B network-contact fluctuation.

**$NH_3$ loading first maximizes shell mixing and then increases the $BH_4^-/NH_3$ network persistence**

The MetaD simulations indicate that Li-shell rearrangement is coupled to changes in the N···B network-contact variable. Unbiased trajectories were therefore analyzed to determine how $NH_3$ loading reorganizes the mixed $BH_4^-/NH_3$ network surrounding the Li coordination shell. $BH_4^-/NH_3$ network edges were defined using the first-shell N···B criterion (**Supplementary Fig. 17**). Ammine borohydrides commonly contain N–H$^{\delta+}$···H$^{\delta-}$–B dihydrogen contacts between protic N-bound and hydridic B-bound H atoms.[16, 17, 19] For the fixed $x$ = 2 model, 38.6% of N···B-associated pairs have minimum H···H < 2.4 Å (**Supplementary Fig. 17b–d**).

The N···B RDFs retain a pronounced first-shell peak across the full composition series,

indicating persistent local association between $NH_3$ and $BH_4^-$ (**Fig. 4a**). However, the first-shell N···B coordination number decreases with increasing $x$, consistent with the decreasing $BH_4^-/NH_3$ ratio as the $NH_3$ fraction increases (inset of **Fig. 4a**). This reduction in local coordination does not imply fragmentation of the contact network. The largest connected-component fraction ($f_{LCC}$) increases sharply with $NH_3$ loading and exceeds 0.9 near $x$ = 2 (**Fig. 4b**), showing that the intermediate composition already contains an extended $BH_4^-/NH_3$ network (**Supplementary Fig. 18**). Increasing $NH_3$ loading therefore produces a network that is less locally dense but more globally connected.

The short-time edge-turnover rate ($k_{turn}$) decreases with increasing $x$ and remains at lower values at $x$ = 3 and $x$ = 4 (**Fig. 4b** and **Supplementary Fig. 19a**), consistent with slower decay of edge persistence at higher $NH_3$ loading (**Supplementary Fig. 19b**). Low-$x$ compositions therefore renew $BH_4^-/NH_3$ contacts more rapidly, whereas $NH_3$-rich compositions retain a larger fraction of their initial edges over the same lag time. Further increasing $NH_3$ loading consequently slows network renewal and increases the temporal persistence of the $NH_3$-rich contact network. Importantly, $k_{turn}$ describes the replacement of N···B contact edges rather than bulk molecular mobility. The faster turnover at low x therefore reflects fluctuations within a sparse network, whereas $x$ = 2 combines extensive connectivity with strong coordination-shell mixing and continued network reorganization.

The descriptor map in **Fig. 4c** summarizes the coupled response of the Li coordination shell and the surrounding $BH_4^-/NH_3$ network. The Li-shell $NH_3$ fraction, $p_N$, increases monotonically with $NH_3$ loading, whereas the shell-mixing descriptor, Φ, reaches a maximum near $x$ = 2. In parallel, the normalized N···B coordination number decreases, $f_{LCC}$ increases and remains high beyond $x$ = 2, and the normalized $k_{turn}$ decreases at higher $NH_3$ loading. Thus, $x$ = 2 combines maximal shell mixing with an extended but dynamically renewing network, whereas $x$ = 3 and 4 exhibit Li–N-dominant coordination together with slower network renewal. These trends identify $x$ = 2 as a highly mixed and dynamically reconfigurable regime, consistent with enhanced configurational disorder and liquid-like mobility.

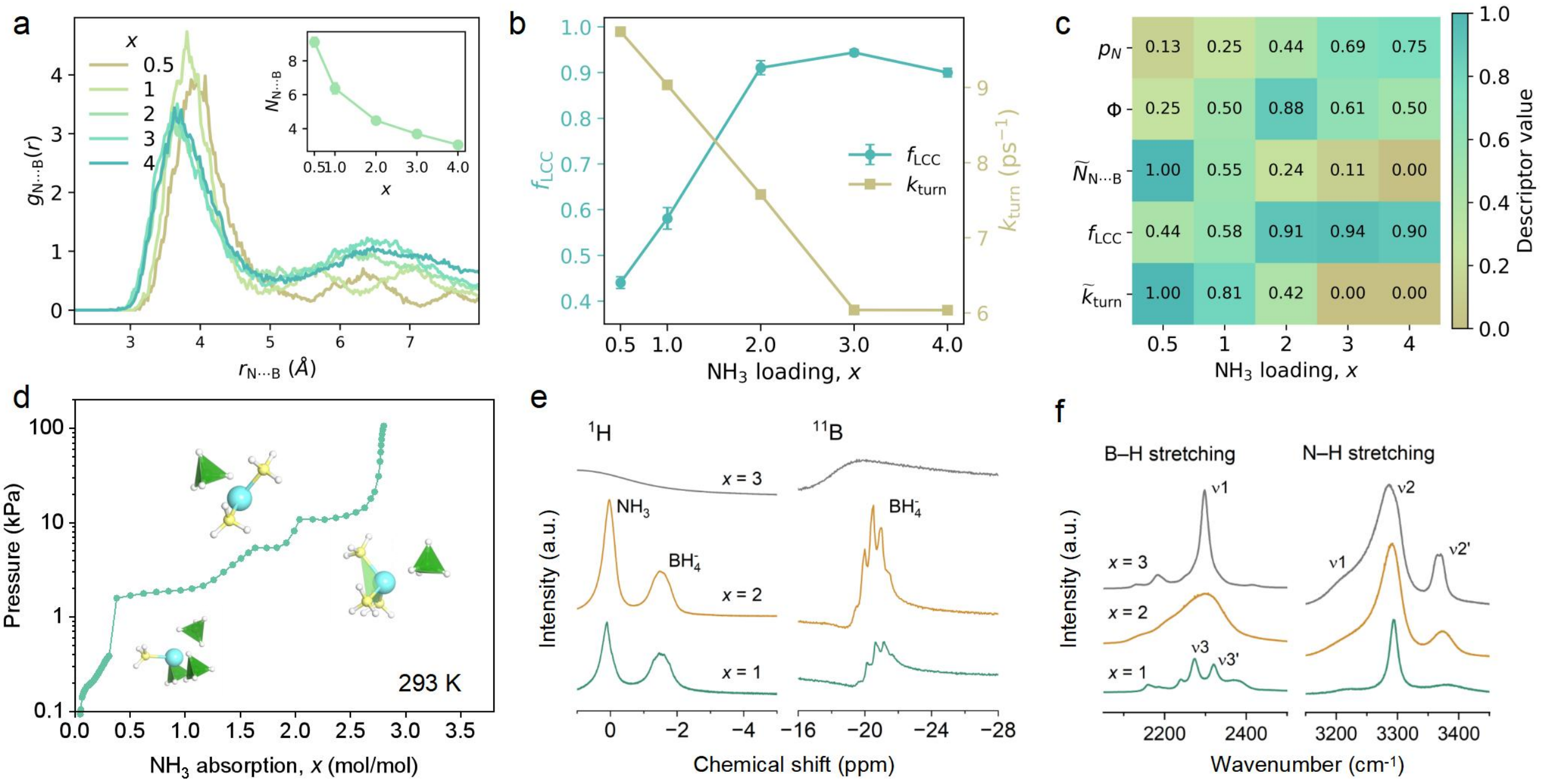

**Fig. 4. $BH_4^-$/$NH_3$ network reorganization in $LiBH_4 \cdot xNH_3$.** **a**, N···B RDFs. Inset, first-shell N···B neighbor number versus $NH_3$ loading. **b**, $NH_3$-loading dependence of the largest connected-component fraction, $f_{LCC}$, and edge-turnover rate, $k_{turn}$. **c**, Descriptor map of $p_N$, Φ, $\tilde{N}_{N\cdots B}$, $f_{LCC}$, and $\tilde{k}_{turn}$. The tilde denotes normalization across the investigated compositions. **d**, Pressure–composition isotherm for $NH_3$ absorption by $LiBH_4$. Inset, representative structures of $LiBH_4 \cdot xNH_3$; cyan, green, yellow, and white spheres represent Li, B, N, and H atoms, respectively. **e**, $^1$H and $^{11}$B NMR spectra of $LiBH_4$ at increasing $NH_3$ absorption amount. **f**, Raman spectra of $LiBH_4$ during $NH_3$ absorption. **Panels d-f** were adapted from Ref.[15].

Room-temperature pressure–composition isotherm (PCI), nuclear magnetic resonance (NMR), and Raman measurements provide experimental tests of this composition-dependent sequence. At 293 K, the sequential formation of $LiBH_4 \cdot NH_3$, $LiBH_4 \cdot 2NH_3$, and $LiBH_4 \cdot 3NH_3$ occurs at approximately 1.83, 5.40, and 10.8 kPa, with corresponding entropy changes of −106, −168, and −98.9 J mol$^{-1}$ K$^{-1}$, respectively (**Fig. 4d**). The substantially larger magnitude of the entropy change for the $x = 1 \rightarrow 2$ uptake step occurs in the same composition range as the maximum in Φ, consistent with pronounced reorganization of the local Li coordination environment near $x \approx 2$. Together with the sloping PCI region between $x \approx 1.5$ and 2 and the reported partial liquefaction,[15, 20, 21, 22] these observations identify an intermediate region associated with partial liquefaction near $x \approx 2$. Clear

Li–N dominance is established only by $x = 3$, coinciding with the recovery of a more rigid and structurally defined ammoniate state. Representative photographs show rigid samples at low and high $NH_3$ absorption levels and a partially liquefied intermediate (**Supplementary Fig. 20**). A video recording directly captures the macroscopic solid–liquid–solid evolution during $NH_3$ desorption, with the instantaneous $NH_3$ pressure displayed throughout (**Supplementary Video 3**).

The relatively sharp $^{1}H$ and $^{11}B$ resonances at $x \approx 2$ are consistent with motional narrowing and a liquid-like local environment (**Fig. 4e**). Quantitative $^{1}H$ and $^{11}B$ linewidth analysis shows relatively narrow resonances in the intermediate absorption region, followed by pronounced broadening at higher $NH_3$ absorption (**Supplementary Fig. 21**). This behavior is consistent with enhanced local motion in the intermediate regime and increased rigidity at higher absorption, without requiring a homogeneous $x = 2$ liquid phase. At $x \approx 3$, the strong attenuation or disappearance of these resonances is consistent with static broadening and the recovery of a rigid $LiBH_4{\cdot}3NH_3$ environment, in good agreement with Li–N-dominant coordination and slower $BH_4^{-}/NH_3$ network renewal. Upon formation of $LiBH_4{\cdot}NH_3$, the B–H stretching band near 2298 $cm^{-1}$ (ν1) is suppressed, whereas the band near 2274 $cm^{-1}$ (ν3) and its adjacent shoulder (ν3') increase in intensity and shift in frequency (**Fig. 4f**). Continued evolution of the B–H and N–H regions from $x = 1$ to 3 further supports the reorganization of the local $BH_4^{-}$ and $NH_3$ environments across the same coordination sequence. Taken together, these measurements support a re-entrant state response governed by the coupled evolution of Li-shell composition, coordination dynamics, and network persistence rather than a simple monotonic progression of $NH_3$ solvation. The liquid-like intermediate nevertheless retains substantial molecular character rather than behaving as a fully dissociated ionic solution, with intact $BH_4^{-}$ and $NH_3$ participating in a dynamically reorganizing coordination network.

## Discussion

The atomistic structures of $LiBH_4{\cdot}2NH_3$, $LiBH_4{\cdot}3NH_3$, and $LiBH_4{\cdot}4NH_3$ remain experimentally unresolved. Therefore, we employed USPEX crystal-structure searches to obtain low-energy, periodically ordered candidates as structural reference states. The 0 K stability analysis places the liquid-like $x$ = 2 composition above the tie line connecting the neighboring $x$ = 1 and $x$ = 3 ammoniates, indicating that its finite-temperature liquid-like behavior cannot be explained by static electronic energies alone. Despite the ordered starting structures, finite-temperature dynamics reveal the greatest degree of disorder at $x$ = 2, while MetaD sampling identifies the broadest coupled coordination space at the same composition. These results characterize local dynamical behavior accessible to the fixed-composition model, rather than proving a homogeneous bulk liquid phase or excluding eutectic melting.

These discrete reference structures represent selected local environments within the experimentally reported intermediate composition ranges and two-phase fields. Fixed-composition periodic AIMD/MetaD cannot resolve macroscopic solid–liquid separation, a finite-temperature liquidus, or phase fractions. Experimental $x$ in this region denotes overall equilibrated $NH_3$ absorption of the $LiBH_4$; local coordination dynamics may coexist with macroscopic phase coexistence. Here, "partially liquefied" describes a macroscopic sample with rigid and fluid fractions, whereas "liquid-like" describes local dynamical behavior without assigning a homogeneous bulk liquid; "liquid" is reserved for an experimentally established macroscopically fluid state. Structural organization beyond the first Li coordination shell, including higher-order $BH_4^-/NH_3$ associations and longer-range packing, may further influence the precise state boundaries. The present models therefore establish systematic composition-dependent coordination trends, whereas the detailed structures of the liquid-like and coexisting phases await direct experimental resolution.

Known ammine metal borohydrides span chain, layer, framework, molecular and complex-ion structures, depending on cation size, valence and preferred coordination.[16, 17, 18] For example, $Mg(BH_4)_2{\cdot}2NH_3$ contains discrete $Mg(BH_4)_2(NH_3)_2$ units packed into layers through dihydrogen contacts, whereas $Mg(BH_4)_2{\cdot}6NH_3$ contains $[Mg(NH_3)_6]^{2+}$ complex cations and $BH_4^-$ counterions.[19]

These structures provide a conventional coordination-chemistry context for the Li trends, while the applicability of the proposed re-entrant mechanism to other cations remains to be established.

Composition-controlled state switching could decouple the fluidity required for transport or processing from the rigidity required during operation. In hydrogen-storage reactions, transient mobility may improve reactant contact and hydrogen transfer, while subsequent rigidification may suppress segregation and restore structural stability. More broadly, chemically induced transient fluidity may facilitate ion transport and interfacial contact in solid/liquid electrolyte systems and could provide a conceptual route to overcoming kinetic limitations during the formation of hydrogen-rich hydride phases. Related forms of dynamic hydride chemistry are already exploited in ammonia synthesis: metal-imide chemical looping separates nitrogen fixation from hydrogenation,[40] whereas ternary ruthenium complex hydrides couple hydridic hydrogen transfer to associative $NH_3$ formation.[41] In electrochemical manufacturing, electrolyte melt infiltration similarly exploits a liquid processing state followed by solidification to improve electrode filling and interfacial contact.[42] An $NH_3$-responsive hydride could analogously provide fluidity for wetting and pore infiltration, followed by rigidification during operation. Practical use would require control of $NH_3$ chemical potential because pressure and temperature affect absorption and hence physical state.[15, 31] The absorption/desorption PCI measurement at 293 K supports reversible $NH_3$ absorption and desorption with composition-dependent hysteresis (**Supplementary Fig. 22**), while the video documents macroscopic solid−liquid−solid evolution during $NH_3$ desorption (**Supplementary Video 3**). $NH_3$ volatility, electrochemical stability, interfacial compatibility, and cycling reversibility therefore remain practical constraints on the proposed fluidity–rigidity switching. Whether similar switching can be realized in ammonia-free hydrides through substitution, eutectic formation, defects, or nanoconfinement remains an important direction for future study.

## Conclusion

In summary, we have identified the interesting mechanism of the long-lasting phenomenon in hydride/ammonia science: $LiBH_4 \cdot xNH_3$ becomes liquid-like not simply as the $NH_3$ content increases, but when Li–N and neighboring $BH_4^-$-derived interactions are most strongly mixed near $x \approx 2$. This mixed-shell regime exhibits the broadest sampled Li–N/N···B coordination region among the simulated compositions, consistent with the experimentally observed liquid-like intermediate. Further ammonia coordination establishes Li–N-dominant coordination and slows $BH_4^-/NH_3$ contact renewal; the resulting increase in network persistence accompanies re-rigidification. Photographs and a desorption video document the re-entrant macroscopic state evolution, while PCI characterizes $NH_3$ absorption and desorption. Quantitative $^1H$ and $^{11}B$ NMR support enhanced local motion near $x = 2$ and recovery of rigidity near $x = 3$, and Raman measurements track the associated local reorganization. Together with the cross-database analysis, these findings identify coordination-shell mixing and network persistence as key determinants of ligand-controlled fluidity and rigidity. This interesting principle may enable transient fluidity for transport and processing, followed by rigidification for structural stability, although its transferability to ammonia-free complex hydrides remains to be tested.

**Acknowledgements**

The authors acknowledge the Center for Computational Materials Science, Institute for Materials Research, Tohoku University for the use of MASAMUNE-IMR (Nos. 202512-SCKXX-0205 and 202512-SCKXX-0218), and the Institute for Solid State Physics (ISSP) at the University of Tokyo for the computational resources.

**Funding**

This work was supported by JSPS KAKENHI (Grant Nos. JP25H01508 and JP25K17845) and the JST Strategic International Collaborative Research Program (SICORP; Grant No. JPMJSC25E2).

**Author Contributions**

Q.W. performed the computational analyses, developed the AI-assisted workflow, analysed the data, prepared the figures, and wrote the original draft. Z.X. performed the experimental investigations, analysed the experimental data, and contributed to the original draft. R.S. contributed to the molecular dynamics calculations, computational analysis, and revision of the manuscript. Y.L. contributed to the development and refinement of the AI agent workflow. H.M. and T.I. contributed to the experimental design and interpretation of the ammonia absorption measurements. E.J.C. and S.O. contributed to the interpretation of the results and revision of the manuscript. F.G. and H.L. supervised the research, acquired funding, and contributed to the interpretation of the work. H.L. originated the central scientific concept, formulated the overall research strategy, and provided the overall scientific leadership throughout the study. All authors discussed the results and reviewed the manuscript.

**Data Availability**

The data supporting the findings of this work are included in the article and its **Supplementary Information**. Literature-derived experimental and computational records analyzed in this work are available through the Digital Battery platform (*DigBat*: https://www.digbat.org) and Digital Hydrogen platform (*DigHyd*: https://www.dighyd.org). Demonstrations of the agent interface and code-assisted analysis are provided in **Supplementary Videos 1** and **2**. A video recording of the macroscopic solid–liquid–solid evolution of $LiBH_4$ during $NH_3$ desorption is provided in **Supplementary Video 3**. Additional data are available from the corresponding authors upon reasonable request.

## Code Availability

Additional custom analysis scripts are available from the corresponding authors upon reasonable request.

## Competing Interests

The authors declare no competing interests.